\documentstyle[12pt]{article}

\def\singlespace {\smallskipamount=3.75pt plus1pt minus1pt
                  \medskipamount=7.5pt plus2pt minus2pt
                  \bigskipamount=15pt plus4pt minus4pt
                  \normalbaselineskip=15pt plus0pt minus0pt
                  \normallineskip=1pt
                  \normallineskiplimit=0pt
                  \jot=3.75pt
                  {\def\smallskip {\vskip\smallskipamount}}
                  {\def\medskip   {\vskip\medskipamount}}
                  {\def\bigskip   {\vskip\bigskipamount}}
                  {\setbox\strutbox=\hbox{\vrule 
                    height10.5pt depth4.5pt width 0pt}}
                  \parskip 7.5pt
                  \normalbaselines}
\def\middlespace {\smallskipamount=5.625pt plus1.5pt minus1.5pt
                  \medskipamount=11.25pt plus3pt minus3pt
                  \bigskipamount=22.5pt plus6pt minus6pt
                  \normalbaselineskip=22.5pt plus0pt minus0pt
                  \normallineskip=1pt
                  \normallineskiplimit=0pt
                  \jot=5.625pt
                  {\def\smallskip {\vskip\smallskipamount}}
                  {\def\medskip   {\vskip\medskipamount}}
                  {\def\bigskip   {\vskip\bigskipamount}}
                  {\setbox\strutbox=\hbox{\vrule 
                    height15.75pt depth6.75pt width 0pt}}
                  \parskip 11.25pt
                  \normalbaselines}
\def\doublespace {\smallskipamount=7.5pt plus2pt minus2pt
                  \medskipamount=15pt plus4pt minus4pt
                  \bigskipamount=30pt plus8pt minus8pt
                  \normalbaselineskip=30pt plus0pt minus0pt
                  \normallineskip=2pt
                  \normallineskiplimit=0pt
                  \jot=7.5pt
                  {\def\smallskip {\vskip\smallskipamount}}
                  {\def\medskip   {\vskip\medskipamount}}
                  {\def\bigskip   {\vskip\bigskipamount}}
                  {\setbox\strutbox=\hbox{\vrule 
                    height21.0pt depth9.0pt width 0pt}}
                  \parskip 15.0pt
                  \normalbaselines}

\begin{document}

\begin{center}
{\bf {\Large A Simple Derivation of the Naked Singularity}}

{\bf {\Large in Spherical Dust Collapse}}

\bigskip\ 

{\bf {Sukratu Barve}\footnote{{e-mail address:
sukkoo@relativity.tifr.res.in}}, T. P. Singh\footnote{{ e-mail address:
tpsingh@tifr.res.in}}}

{\it Tata Institute of Fundamental Research,}

{\it Homi Bhabha Road, Mumbai 400 005, India.}

\smallskip\ 

{\bf Cenalo Vaz\footnote{{ e-mail address: cvaz@haar.pha.jhu.edu}}%
\footnote{{ On leave of absence from the Universidade do Algarve, Faro,
Portugal}}}

{\it Department of Physics, The Johns Hopkins University, }

{\it Baltimore, MD 21218, USA}

\smallskip\ 

{\bf Louis Witten\footnote{{ e-mail address: witten@physics.uc.edu}}}

{\it Department of Physics, University of Cincinnati,}

{\it Cincinnati, OH 45221-0011, USA}

\bigskip\ 

{\bf Abstract}
\end{center}

\noindent We describe a simple method of determining whether the singularity 
that forms in the spherically symmetric collapse of inhomogeneous dust is
naked or covered. This derivation considerably simplifies the analysis given 
in the earlier literature, while giving the same results as have been 
obtained before.

\middlespace

\newpage\ \ \noindent Various authors have shown that the spherical
gravitational collapse of inhomogeneous dust results in the formation of a
curvature singularity which is naked for certain initial conditions, and
covered for other initial conditions. Probably the first (numerical) results
on this problem are due to Eardley and Smarr \cite{es} . This was followed
by the analytical work of Christodoulou \cite{cd} and Newman \cite{ne} who
considered smooth initial data. Their results were generalised, among
others, by Ori and Piran \cite{op}, and by Dwivedi, Jhingan, Joshi and Singh 
\cite{o}, \cite{sj}. As is only to be expected, in one way or the other,
these works all deal with propagation of null geodesics in the spacetime of
collapsing dust.

While successive works have succeeded in simplifying the earlier analysis,
perhaps it can be said that the discussions continue to remain somewhat
involved. In the present paper, we describe a short but straightforward 
method of showing whether the naked singularity in the dust model is covered or
naked. We reproduce results obtained previously by other methods.
We restrict attention to the marginally bound dust collapse - similar 
principles may be used to derive results for the non-marginally bound case.

In comoving coordinates $(t,r,\theta ,\phi )$ the spacetime metric for
spherical dust collapse is given by 
\begin{equation}
\label{met}ds^2=dt^2-R^{\prime 2}dr^2-R^2d\Omega ^2 
\end{equation}
where $R(t,r)$ is the area radius at time $t$ of the shell having the
comoving coordinate $r$. A prime denotes partial derivative w.r.t. $r$. The
energy-momentum tensor for dust has only one non-zero component $%
T_0^0=\epsilon (t,r)$, which is the energy density. The Einstein equations
for the collapsing cloud are 
\begin{equation}
\label{e1}\frac{8\pi G}{c^4}\epsilon (t,r)=\frac{F^{\prime }}{R^2R^{\prime }}%
,\;\;\dot R^2=\frac{F(r)}R. 
\end{equation}
A dot denotes partial derivative w.r.t. time $t$. The function $F(r)$
results from the integration of the second order equations. Henceforth we
shall set $8\pi G/c^4=1$.

The second of these equations can be easily solved to get 
\begin{equation}
\label{R}R^{3/2}(t,r)=r^{3/2}-\frac 32\sqrt{F}t 
\end{equation}
where we have used the freedom in the scaling of the comoving coordinate $r$
to set $R(0,r)=r$ at the starting epoch of collapse, $t=0$. It follows from
the first equation in (\ref{e1}) that the function $F(r)$ gets fixed once
the initial density distribution $\epsilon (0,r)=\rho (r)$ is given, i.e. 
\begin{equation}
\label{eff}F(r)=\int \rho (r)r^2dr. 
\end{equation}
Hence $F(r)$ has the interpretation of being twice the mass to the interior
of the shell labeled $r$. If the initial density $\rho (r)$ has a series
expansion 
\begin{equation}
\label{de}\rho (r)=\rho _0+\rho _1r+\frac 1{2!}\rho _2r^2+\frac 1{3!}\rho
_3r^3+... 
\end{equation}
near the center $r=0$, the resulting series expansion for the mass function $%
F(r)$ is 
\begin{equation}
\label{ma}F(r)=F_0r^3+F_1r^4+F_2r^5+F_3r^6+... 
\end{equation}
where $F_q=\rho _q/q!(q+3)$, and $q=0,1,2,3..$We note that we could set $%
\rho _1=0$ without in any way affecting the conclusions of this paper.
Further, the first non-vanishing derivative in the series expansion in (\ref
{de}) should be negative, as we will consider only density functions which
decrease as one moves out from the center.

According to (\ref{R}) the area radius of the shell $r$ shrinks to zero at
the time $t_c(r)$ given by 
\begin{equation}
\label{tc}t_c(r)=\frac{2r^{3/2}}{3\sqrt{F(r)}}. 
\end{equation}
At $t=t_c(r)$ the Kretschmann scalar 
\begin{equation}
\label{kr}K=12\frac{F^{\prime 2}}{R^4R^{\prime 2}}-32\frac{FF^{\prime }}{%
R^5R^{\prime }}+48\frac{F^2}{R^6} 
\end{equation}
diverges at the shell labeled $r$ and hence this represents the formation of
a curvature singularity at $r.$ In particular, the central singularity, i.e.
the one at $r=0$, forms at the time 
\begin{equation}
\label{sing}t_0=\frac 2{3\sqrt{F_0}}=\frac 2{\sqrt{3\rho _0}}. 
\end{equation}
At $t=t_0$ the Kretschmann scalar diverges at $r=0$. Near $r=0,$ we can
expand $F(r)$ and approximately write for the singularity curve 
\begin{equation}
\label{texp}t_c(r)=t_0-\frac{F_n}{3F_0^{3/2}}r^n. 
\end{equation}
Here, $F_n$ is the first non-vanishing term beyond $F_0$ in the expansion (%
\ref{ma}). We note that $t_c(r)>t_0$, since $F_n$ is negative.

We wish to investigate if the singularity at $t=t_0,r=0$ is naked, i.e. are
there one or more outgoing null geodesics which terminate in the past at the
central singularity. We restrict attention to radial null geodesics. Let us
start by assuming that one or more such geodesics exist, and then checking
if this assumption is correct. Let us take the geodesic to have the form 
\begin{equation}
\label{geo}t=t_0+ar^\alpha 
\end{equation}
to leading order, in the $t-r$ plane, where $a>0$, $\alpha >0$. In order for
this geodesic to lie in the spacetime, we conclude by comparing with (\ref
{texp}) that $\alpha \geq n$, and in addition, if $\alpha =n$, then $%
a<-F_n/3F_0^{3/2}$.

As is evident from the form (\ref{met}) of the metric, an outgoing null
geodesic must satisfy the equation 
\begin{equation}
\label{ng}\frac{dt}{dr}=R^{\prime }. 
\end{equation}
In order to calculate $R^{\prime }$ near $r=0$ we first write the solution (%
\ref{R}) with only the leading term $F_n$ retained in $F(r)$ in (\ref{ma}).
This gives 
\begin{equation}
\label{Ra}R=r\left( 1-\frac 32\sqrt{F_0}\left[ 1+\frac{F_n}{2F_0}r^n\right]
\,t\right) ^{2/3}. 
\end{equation}
Differentiating this w.r.t. $r$ gives 
\begin{equation}
\label{Rp}R^{\prime }=\left( 1-\frac 32\sqrt{F_0}\left[ 1+\frac{F_n}{2F_0}%
r^n\right] \,t\right) ^{-1/3}\,\,\left( 1-\frac 32\sqrt{F_0}t-\frac{(2n+3)F_n%
}{4\sqrt{F_0}}r^nt\right) . 
\end{equation}

Along the assumed geodesic, $t$ is given by (\ref{geo}). Substituting this
in $R^{\prime }$ and equating the resulting $R^{\prime }$ to $dt/dr=\alpha
ar^{\alpha -1}$ gives 
\begin{equation}
\label{maj}\alpha ar^{\alpha -1}=\frac{\,\left( 1-\frac 32\sqrt{F_0}\left[
t_0+ar^\alpha \right] -\frac{(2n+3)F_n}{4\sqrt{F_0}}r^n\left[ t_0+ar^\alpha
\right] \right) }{\left( 1-\frac 32\sqrt{F_0}\left[ 1+\frac{F_n}{2F_0}%
r^n\right] \,\left[ t_0+ar^\alpha \right] \right) ^{1/3}}\,. 
\end{equation}
$\,$ This is the key equation. If it admits a self-consistent solution then
the singularity will be naked (i.e. at least one outgoing null geodesic will
terminate at the singularity), otherwise not. We simplify this equation by
putting in the requirement mentioned earlier, that $\alpha \geq n$. Consider
first $\alpha >n$. In this case we get, to leading order 
\begin{equation}
\label{oo}\alpha ar^{\alpha -1}=\left( 1+\frac{2n}3\right) \left( -\frac{F_n%
}{2F_0}\right) ^{2/3}r^{2n/3} 
\end{equation}
which implies that $\alpha =1+2n/3$, and $a=(-F_n/2F_0)^{2/3}$. By
substituting integral values for $n$ we find that only for $n=1$ and $n=2$
the condition $\alpha >n$ is satisfied. Hence the singularity is naked for $%
n=1$ and $n=2$, i.e. for the models $\rho _1<0$ and for $\rho _1=0,\rho _2<0 
$. There is at least one outgoing geodesic given by (\ref{geo}) which
terminates in the central singularity in the past. If $n>3$ then the
condition $\alpha >n$ cannot be satisfied and the singularity is not naked.
This is the case $\rho _1=\rho _2=\rho _3=0$.

Consider next that $\alpha =n$. In this case we get from (\ref{maj}) that 
\begin{equation}
\label{n3}nar^{n-1}=\frac{-\frac 32a\sqrt{F_0}-\frac{(2n+3)F_n}{6\sqrt{F_0}}%
}{\left( -\frac{F_n}{2F_0}-\frac{3a}2\sqrt{F_0}\right) ^{1/3}}r^{2n/3} 
\end{equation}
which implies that $n=3$ and gives an implicit expression for $a$ in terms
of $F_3$ and $F_0$. This expression for $a$ can be simplified to get the
following quartic for $a$: 
\begin{equation}
\label{qu3}12\sqrt{F_0}%
a^4-a^3(-4F_3/F_0+F_0^{3/2})-3F_3a^2-3F_3^2/F_0^{3/2}a-(F_3/F_0)^3=0. 
\end{equation}
By defining $b=a/F_0$ and $\xi =F_3/F_0^{5/2}$ this quartic can be written
as 
\begin{equation}
\label{qu4}4b^3(3b+\xi )-(b+\xi )^3=0. 
\end{equation}
The singularity will be naked if this equation admits one or more positive
roots for $b$ which satisfy the constraint $b<-\xi /3$. This last inequality
is the same as the condition $a<-F_n/3F_0^{3/2}$ given below equation (\ref
{geo}). We note that $\xi $ is negative. This quartic can be made amenable
to further analysis by substituting $Y=-2b/\xi ,$ and then $\eta
=-1/6\xi $, so as to get 
\begin{equation}
\label{qn2}Y^3(Y-2/3)-\eta (Y-2)^3=0. 
\end{equation}
As discussed in \cite{sj} this quartic has two positive real roots provided $%
\eta \geq \eta _1$ or $\eta \leq \eta _2$ where 
\begin{equation}
\label{al}\eta _1=\frac{26}3+5\sqrt{3},\;\eta _2=\frac{26}3-5\sqrt{3}.\; 
\end{equation}
We also require that $Y<2/3$. By examining the quartic (\ref{qn2}) one can
see that if $\eta \geq \eta _1$ then $Y\geq 2$; hence this range of $\eta $
is ruled out. Thus the singularity is naked provided $\eta \leq \eta _2$, or
equivalently $\xi \leq -25.9904$.

This completes the analysis to decide whether or not the central singularity
is naked, and we get the same results as have been given earlier in the
literature, albeit in a much simpler manner. Now we examine whether or not
there is an entire family of radial null geodesics which terminate at the
naked singularity. For this purpose we assume a solution for the geodesics
correct to one order beyond the solution (\ref{geo}), i.e. we take 
\begin{equation}
\label{geo2}t=t_0+ar^\alpha +dr^{\alpha +\beta }\,. 
\end{equation}
where $d$ and $\beta $ are constants to be determined, and $a$ and $\alpha $
take the values calculated above. As before, we substitute this form of $%
t(r) $ in the expression (\ref{Rp}) for $R^{\prime }$ to get 
\begin{equation}
\label{Rp2}R^{\prime }=\frac{\,\left( 1-\frac 32\sqrt{F_0}\left[
t_0+ar^\alpha +dr^{\alpha +\beta }\,\right] -\frac{(2n+3)F_n}{4\sqrt{F_0}}%
r^n\left[ t_0+ar^\alpha +dr^{\alpha +\beta }\,\right] \right) }{\left(
1-\frac 32\sqrt{F_0}\left[ 1+\frac{F_n}{2F_0}r^n\right] \,\left[
t_0+ar^\alpha +dr^{\alpha +\beta }\,\right] \right) ^{1/3}}\,. 
\end{equation}
Next, we equate this $R^{\prime }$ to the $dt/dr$ calculated from (\ref{geo2}%
). For the cases $n=1,2$ we get, after retaining terms up to second order 
\begin{equation}
\label{waa}\alpha ar^{\alpha -1}+(\alpha +\beta )dr^{\alpha +\beta
-1}=\left( 1+\frac{2n}3\right) \left( -\frac{F_n}{2F_0}\right)
^{2/3}r^{2n/3}+Dr^{\alpha -n/3}. 
\end{equation}
As before, at the leading order $a$ and $\alpha $ get fixed. At the next
order, we get $\beta =1+n/3$ and $d=D/(2+n/3)$ where 
\begin{equation}
\label{dee}D=\frac 32\sqrt{F_0}\left( -\frac{F_n}{2F_0}\right) ^{1/3}\left[
-1+\frac 13\left( 1+\frac{2n}3\right) \left( -\frac{F_n}{2F_0}\right)
^{-2/3}\right] . 
\end{equation}
It thus follows, according to (\ref{geo2}), that when $n=1,2$ there is to this
order only one outgoing geodesic, having the values of $d$ and $\beta $ given 
above.

Consider next $n=3$. By repeating the above calculation of $R^{\prime }$ we
get 
\begin{equation}
\label{rag}R^{\prime }=\frac 322^{1/3}F_0\frac{\xi +b}{\left( \xi +3b\right)
^{1/3}}r^2+3.2^{1/3}bd\frac 1{\left( \xi +3b\right) ^{4/3}}r^{2+\beta
}+O(r^5).
\end{equation}
Here, $O(r^5)$ is a term of order $r^5$ which is independent of $d$. Further
analysis depends on whether or not $\beta $ is less than $3$. Assume first
that $\beta <3$. Then the $O(r^5)$ term can be ignored. By equating $%
R^{\prime }$ to $dt/dr=3ar^2+(3+\beta )dr^{2+\beta }$, we get the earlier
quartic (\ref{qu4}) for $b$. At the next order, $d$ drops out and one gets
an equation for $\beta $, i.e. 
\begin{equation}
\label{eff2}3+\beta =3b.2^{1/3}\frac 1{\left( \xi +3b\right) ^{4/3}}.
\end{equation}

Since $d$ drops out, this means that $d$ is arbitrary, and there will be an
entire family of outgoing null geodesics terminating at the singularity,
provided $\beta $ is non-negative. It is essential that $\beta $ be
non-negative, otherwise these geodesics will not lie in the spacetime, as is
evident from a comparison with the singularity curve (\ref{texp}). As we saw
above, the quartic (\ref{qn2}) and hence (\ref{qu4}) has two positive roots
in the naked singular range. We now show that $\beta $ as defined in (\ref
{eff2}) is positive at one of the roots, and negative at the other root. Let
us write the quartic (\ref{qn2}) as $V(Y)=0$ where 
\begin{equation}
\label{poly}V(Y)=Y^3\left( \frac 23-Y\right) -\alpha (2-Y)^3.
\end{equation}
It is then easily shown that 
\begin{equation}
\label{wow2}\left( \frac{dV}{dY}\right) _{Y=Y_0}=-\beta Y_0^2\left( \frac
23-Y_0\right) 
\end{equation}
where $Y_0$ is a real, positive root of the quartic. Since the derivative $%
dV/dY$ must be positive at one of the roots and negative at the other, and
since $Y_0<2/3$ it follows that $\beta $ is positive at one of the roots and
negative at the other. Hence one of the roots admits only one outgoing
geodesic (for which $d$=$0$) while the other root admits an entire family of
outgoing geodesics. It is easily verified that the family emerges from the
larger of the two roots, which lies closer to the singularity curve (\ref
{texp}).

However, it also turns out, as is verified numerically, that the positive
value of $\beta $ does not remain below $3$ for all $\xi $. It can be shown
that $\beta $ can be written as 
\begin{equation}
\label{be}\beta =-\frac{Y^2-8Y+4}{\left( 2/3-Y\right) \left( 2-Y\right) }. 
\end{equation}
For values of $\xi $ smaller than a certain critical value, $\beta $ becomes
larger than $3$, so that then the $O(r^5)$ term in (\ref{rag}) dominates
over the term proportional to $r^{2+\beta }.$ For such cases, we get from $%
dt/dr=R^{\prime }$ that $\beta =3$, and $d$ also gets fixed at a particular
value. In order to see the family of outgoing rays we will have to look at
higher order terms in the various expansions.

Consider next the case of ingoing rays, given by $dt/dr=-R^{\prime }$, for
which we take 
\begin{equation}
\label{in}t=t_0-er^3-gr^{3+\gamma } 
\end{equation}
(We consider only the case $n=3$). The expression for $R^{\prime }$ is 
\begin{equation}
\label{rp2}R^{\prime }=\frac 322^{1/3}F_0\frac{h-\xi }{\left( 3h-\xi \right)
^{1/3}}r^2+\frac{3hg}42^{4/3}\frac 1{\left( \xi -3h\right)
^{4/3}}r^{2+\gamma } 
\end{equation}
where $h=eF_0$. Equating this to $dt/dr=-3er^2-(3+\gamma )gr^{2+\gamma }$
gives, at order $r^2,$ the following quartic for $h$%
\begin{equation}
\label{ing}h^3(12h-4\xi )+(h-\xi )^3=0 
\end{equation}
which admits a positive root $h$ for all $\xi $, as expected, so that there
is always an ingoing ray. However at the next order, we get the relation 
\begin{equation}
\label{no}-(3+\gamma )=\frac{3h}42^{4/3}\frac 1{\left( \xi -3h\right)
^{4/3}} 
\end{equation}
which cannot be satisfied, unless $g=0$, since the l.h.s. is negative and
the r.h.s. positive. This shows that there is only one ingoing ray to all 
orders in the expansion.

\bigskip\ 

{\noindent {\bf ACKNOWLEDGMENTS}}

\smallskip

\noindent We acknowledge partial support of the Junta Nacional de
Investigac\~ao Cient\'ifica e Tecnol\'ogica (JNICT) Portugal, under contract
number CERN/S/FAE/1172/97. C. V. and L. W. acknowledge the partial
support of NATO, under contract number CRG 920096; L. W. acknowledges the
partial support of the U. S. Department of Energy under contract number
DOE-FG02-84ER40153 and C.V. acknowledges the partial support of the FCT under
contract number FMRH/BSAB/54/98.

\end{document}